\documentclass[twocolumn]{aastex63}

\usepackage{graphicx}
\usepackage{dcolumn}
\usepackage{bm}
\usepackage{acronym}
\usepackage[cal=boondox]{mathalfa}
\usepackage{aas_macros}

\acrodef{GW}{gravitational wave}
\acrodef{PE}{Parameter Estimation}
\acrodef{MCMC}{Markov Chain Monte Carlo}
\acrodef{DOF}{degrees of freedom}
\acrodef{AGN}{Active Galactic Nuclei}

\newcommand{\checkme}[1]{{#1}}
\newcommand{\peakonemch}{\checkme{8}}
\newcommand{\peaktwomch}{\checkme{14}}
\newcommand{\peakthreemch}{\checkme{26}}
\newcommand{\peakfourmch}{\checkme{45}}
\newcommand{\peakonem}{\checkme{9}}
\newcommand{\peaktwom}{\checkme{16}}
\newcommand{\peakthreem}{\checkme{30}}
\newcommand{\peakfourm}{\checkme{57}}
\newcommand{\ratem}{\checkme{27.3}}
\newcommand{\rateu}{\checkme{12.9}}
\newcommand{\ratel}{\checkme{9.7}}
\newcommand{\lrate}{\checkme{17.6}}
\newcommand{\rrate}{\checkme{40.2}}
\newcommand{\lmratio}{\checkme{0.61}}
\newcommand{\rmratio}{\checkme{1.0}}
\newcommand{\lspinz}{\checkme{-0.36}}
\newcommand{\rspinz}{\checkme{+0.36}}

\graphicspath{{figures/}}

\begin{document}

\preprint{APS/123-QED}

\title{The Emergence of Structure in the Binary Black Hole Mass Distribution}

\correspondingauthor{Vaibhav Tiwari}
\email{tiwariv@cardiff.ac.uk}

\author[0000-0002-1602-4176]{Vaibhav Tiwari}
\affiliation{Gravity Exploration Institute \\
School of Physics and Astronomy, \\ 
Cardiff University, Queens Buildings, The Parade \\ 
Cardiff CF24 3AA, UK.}
\author[0000-0001-8480-1961]{Stephen Fairhurst}
\affiliation{Gravity Exploration Institute \\
School of Physics and Astronomy, \\ 
Cardiff University, Queens Buildings, The Parade \\ 
Cardiff CF24 3AA, UK.}

\date{\today}

\begin{abstract}
We use the gravitational wave signals from binary black hole merger events observed by LIGO and Virgo to reconstruct the underlying mass and spin distributions of the population of merging black holes.  We reconstruct the population using the mixture model framework VAMANA \citep{vamana} using observations in GWTC-2 occurring during the first two observing runs and the first half of the third run (O1, O2, and O3a).  Our analysis identifies a structure in the chirp mass distribution of the observed population.  Specifically, we identify peaks in the chirp mass distribution at \peakonemch, \peaktwomch, \peakthreemch, and \peakfourmch $M_\odot$ and a complementary structure in the component mass distribution with an excess of black holes at masses of \peakonem, \peaktwom, \peakthreem\text{ }and \peakfourm $M_{\odot}$. Intriguingly, for both the distributions, the location of subsequent peaks are separated by a factor of around two and there is a lack of mergers with chirp masses of $10-12 M_{\odot}$. The appearance of multiple peaks is a feature of a hierarchical merger scenario when, due to a gap in the black-hole mass spectrum, a pile-up occurs at the first peak followed by mergers of lower mass black-holes to hierarchically produce higher mass black-holes. However, cross-generation merger peaks and observations with high spins are also predicted to occur in such a scenario that we are not currently observing. The results presented are limited in measurement accuracy due to small numbers of observations but if corroborated by future gravitational wave observations these features have far-reaching implications.
\end{abstract}


\section{Introduction} \label{sec:intro}

The field of \ac{GW} astronomy is flourishing. The current tally of confidently detected \ac{GW} signals now stands at around 50 \citep{2019PhRvX...9c1040A, 2019ApJ...872..195N, 2020ApJ...891..123N, 2019PhRvD.100b3007Z, 2020PhRvD.101h3030V, 2020arXiv201014527A, 2020arXiv201014533T}, the majority of which arise from the merger of black hole binaries.  The first observation, GW150914 \citep{2020PhRvL.125j1102A}, contained black holes more massive than any previously known stellar-mass black holes.  More recently, observations have provided the first evidence for unequal mass binaries \citep{2020PhRvD.102d3015A}, the most massive black hole to date \citep{2020PhRvL.125j1102A} which challenges stellar evolution models, and potentially also the least massive black hole \citep{2020ApJ...896L..44A}.  In addition to these exceptional events, the remaining observed binaries are beginning to reveal the underlying properties of the black hole population in the Universe.

The mass and spin distribution for the population of merging binary black holes are of particular interest. Distribution of these parameters can potentially inform about their formation through stellar evolution in the field or due to dynamical interaction in star clusters; the two environments that have been long accepted as the nurseries for the compact binary formation \citep{2008ApJ...676.1162S}. Although there has been uncertainty on the spin distribution of binary black holes it has been anticipated that the mass distribution will have some semblance to the stellar initial mass function \citep{1955ApJ...121..161S, 2001MNRAS.322..231K}. Inspired by this anticipation there have been attempts at modeling the mass distribution using power-laws \citep{TheLIGOScientific:2016pea, 2017PhRvD..95j3010K, 2019PhRvD.100d3012W, LIGOScientific:2018jsj,2019ApJ...878L...1P} which was extended to include Gaussians to model a possible high-mass cut-off due to pair-instability supernova \citep{2018ApJ...856..173T, LIGOScientific:2018jsj} and massive events beyond this high-mass cut-off proposed to be formed due to hierarchical merger of lighter black holes \citep{2020arXiv201014533T}. 

In this article, we present the reconstructed mass and spin distribution, and the estimated merger rate using the flexible mixture-model framework VAMANA \citep{vamana}.  This framework is distinct from analyses presented in, e.g. \cite{2020arXiv201014533T} in two critical ways.  Firstly, the mass distribution of the population is modeled based upon the observed \textit{chirp mass} and mass ratios of events.  While astrophysical formation scenarios might provide predictions on the primary mass distribution, its measurement is degenerate with the mass ratio of the system.  In contrast, the chirp mass
\begin{equation}\label{eq:mchirp}
    \mathcal{M} = \frac{(m_1\,m_2)^{(3/5)}}{(m_1 + m_2)^{(1/5)}}
\end{equation}
is best measured for the majority of observed systems as this determines the leading-order gravitational wave emission during the inspiral of the two black holes.  The second difference is that we employ a flexible fitting based upon a mixture model of multiple Gaussians, rather than a parametrised fit to the data.  These choices allow us to extract structure from the observed signals, rather than fitting pre-specified models to the data.

We also use the observed binary mergers to infer the distributions of mass ratios and spins aligned with the orbital angular momentum.  Both of these affect the emitted waveform but have a less significant impact than the chirp mass.  Furthermore, there is a well-known degeneracy between the two \citep{Cutler:1994ys, 2013PhRvD..87b4035B}.  The recovered distributions are consistent with those reported in \cite{2020arXiv201014533T}.  In particular, we observe that the aligned spin components are small in magnitude while the mass ratio distribution is clearly peaked towards equal masses, although there is evidence for unequal mass binaries in the population.  The in-plane spins impact the emitted waveform through spin-induced orbital precession \citep{1994PhRvD..49.6274A}, which is, in principle, measurable in the observed signal.  To date \citep{2020arXiv201014527A, 2020PhRvD.102d3015A}, there is no strong evidence for precession in single GW observations, and here we do not consider the possibility of a population of weakly precessing signals.  We note that the existence of precession will have only a weak impact on the inferred masses and aligned spins of the binary.  In addition, we do not model any redshift dependence of the rate and assume that it is constant in comoving volume.  Under this assumption, we estimate the merger rate to be $\ratem^{+\rateu}_{-\ratel}\,\mathrm{Gpc}^{-3} \mathrm{yr}^{-1}$ that is consistent with the merger rate reported in \citep{2020arXiv201014533T} but with a slightly tighter 90\% confidence interval.

The remainder of the paper is laid out as follows:
We briefly discuss the analysis in section \ref{analysis}, the reconstructed population's properties in section \ref{popprop}, the hierarchical structure in the  reconstructed mass distribution \ref{sec:hierarchy} and it's astrophysical implication in section \ref{hiermerge}.  

\section{analysis and selected events}
\label{analysis}

We analyse the binary black-hole observations up to GWTC-2 \citep{2019PhRvX...9c1040A, 2020arXiv201014527A}, made during the first two LIGO and Virgo observing runs (O1 and O2) and the first half of the third run (O3a). Compared to event selection made in \citep{2020arXiv201014533T} we make stricter choices, i.e., we select only observations that passed the false alarm rate threshold of once in ten years in any of the search pipelines GstLAL, PyCBC, and PyCBC-BBH \citep{2020arXiv201014527A}. The selected search analyses focus specifically on the \ac{GW} signals that originate from compact binary coalescence.  The false rate of one event per decade is stricter than imposed in \cite{2020arXiv201014533T}, and is motivated by a desire to have an essentially pure \ac{GW} population, noting that the observations span approximately one year of data.  Due to our choice, we do not include GW170729, GW190424, GW190514, and GW190731 in our analysis. Furthermore, we restrict attention to Black Hole Binaries and therefore exclude the two observed binary neutron star signals (GW170817, GW190425 \citep{2017PhRvL.119p1101A, 2020ApJ...892L...3A}) from the analysis, as well as GW190814 \citep{2020ApJ...896L..44A}, whose less massive component has a mass of $2.6 M_{\odot}$, which lies below the accepted black hole mass distribution \citep{2011ApJ...741..103F, 2012ApJ...757...36K, 2012ApJ...759...52D}.  Finally, we exclude the most massive observation to date GW190521 \citep{2020PhRvL.125j1102A}, as this may be a significant outlier or may have formed from a different formation mechanism than the other observations.
\footnote{We have performed the analysis including GW190521 and GW190814. GW190521 creates a very broad peak at its mean chirp-mass value and GW190814 strengthens the first peak while impacting the component mass distribution. Currently, we are treating these two observations as outliers and the remaining observations as \emph{vanilla} binary black-hole binaries}
Moreover, the inferred chirp-mass of GW190521 heavily depends on the assumptions used in the parameter recovery; allowing for eccentricity in the orbit or for high mass ratios significantly changes the measured chirp-mass \citep{2020arXiv200905461G, 2021ApJ...907L...9N}.%
The number of chosen observations is 39 in total. 

The observed binary population will depend both upon the population of black holes in binaries in the Universe as well as the sensitivity of the gravitational wave network to binaries of particular masses and spins.  In order to infer the astrophysical population, we must accurately model the sensitivity of the gravitational wave detectors and searches over the parameter space.  This is achieved by estimating the sensitivity of the searches towards simulated signals added to the data set \citep{2018CQGra..35n5009T}. We follow an identical procedure in correcting for the selection effects as described in \cite{2020arXiv201014533T}. All the \ac{PE} samples and simulation campaign's data is now publicly available \citep{losc}. Out of various \ac{PE} samples available we have chosen the ones obtained using the waveform model \texttt{IMRPhenomPv2} \citep{2012PhRvD..86j4063S, 2014PhRvL.113o1101H}, primarily because they are available for all our selected events. However, we note that GW190412 has exhibited the presence of higher harmonics and \texttt{IMRPhenomPv2} is not best suited for estimating the parameters of this signal.  While the inclusion of higher harmonics may affect the inferred mass ratio, it will not impact the measured chirp-mass of the source.

We model the population using the flexible population modeler VAMANA \citep{vamana} that uses a mixture-model to reconstructs the chirp mass, aligned spin component, and the mass ratio distribution.%
\footnote{Currently, VAMANA does not model the redshift evolution of the merger rate. We instead fix the redshift evolution to uniform in co-moving using the  Planck15 cosmology \citep{2016A&A...594A..13P}}
Each component of the mixture model is composed of two Gaussians and a power-law to model the chirp mass, aligned spins and the mass ratio distributions respectively. The final density is written as a weighted sum of these components. The weights follow a uniform prior and are proposed using a Dirichlet distribution. The likelihood is extended to include a Poisson term for the estimation of the merger rate \citep{extended_lkl}. The reconstructed distributions are robust for a range of choices on the number of components between 5 and 17. In the asymptotic limit of a large number of observations, we expect the mean of the reconstructed distribution to be very close to the true distribution \citep{judith2011}. Modeling individual observations with a separate Gaussian will obtain the best fit with the data but will increase the complexity of the model.  We have calculated the marginal likelihood for the results with a varying number of components and found that it is roughly constant for models with between 7 and 15 components, with the 9 component model providing the highest likelihood.  For this reason, the results presented here are from the 9 component analysis, but these are robust under changes in the number of Gaussian components as shown in the lower plots in Figure \ref{fig:post_mc}.

We choose prior distributions in the same way as for modeling the population from GWTC-1 in \cite{vamana}. These remain suitable for the current observations, even though the mass range observed has increased and multiple mergers with unequal masses have been observed. The prior distribution for the Gaussian peaks modeling the chirp mass is distributed uniformly in the log of the mass and can acquire values between 5.4 $M_\odot$ and 52.7 $M_\odot$. These values correspond to the minimum chirp mass estimate of the lightest observation and the 80th percentile of the chirp mass estimates for the most massive observation. The power law modeling the mass ratio distribution has an exponent, $\alpha_q$, uniform between -8 and 1, and the minimum mass ratio, $q_{\mathrm{min}}$, is between 0 and 0.95. The prior used for Gaussians modeling the aligned-spin is identical to the ones used in \cite{vamana}.  These are broad choices and, combined with the fact that the final distribution is comprised of the sum of multiple components, are able to accurately represent a broad range of distributions.  To prevent overfitting of the results, we require the posterior for the modeled population to be within a threshold distance measure from the reference population, where we choose $r_{\mathrm{eff}} = 0.2$ in this analysis \cite[please see equation 5 in][]{vamana}.
The reference population is a simple phenomenological fit to the data using a power-law for modeling the chirp mass, a single truncated Gaussian to model the aligned spin, and a single power-law to model the mass ratio.

We refer the interested reader to section IIIB of \cite{vamana} for a detailed description of the model. 


\section{The Mass, Spin Distribution and the Merger Rate}
\label{popprop}

\begin{figure*}
    \centering
    \includegraphics[width=0.95\textwidth]{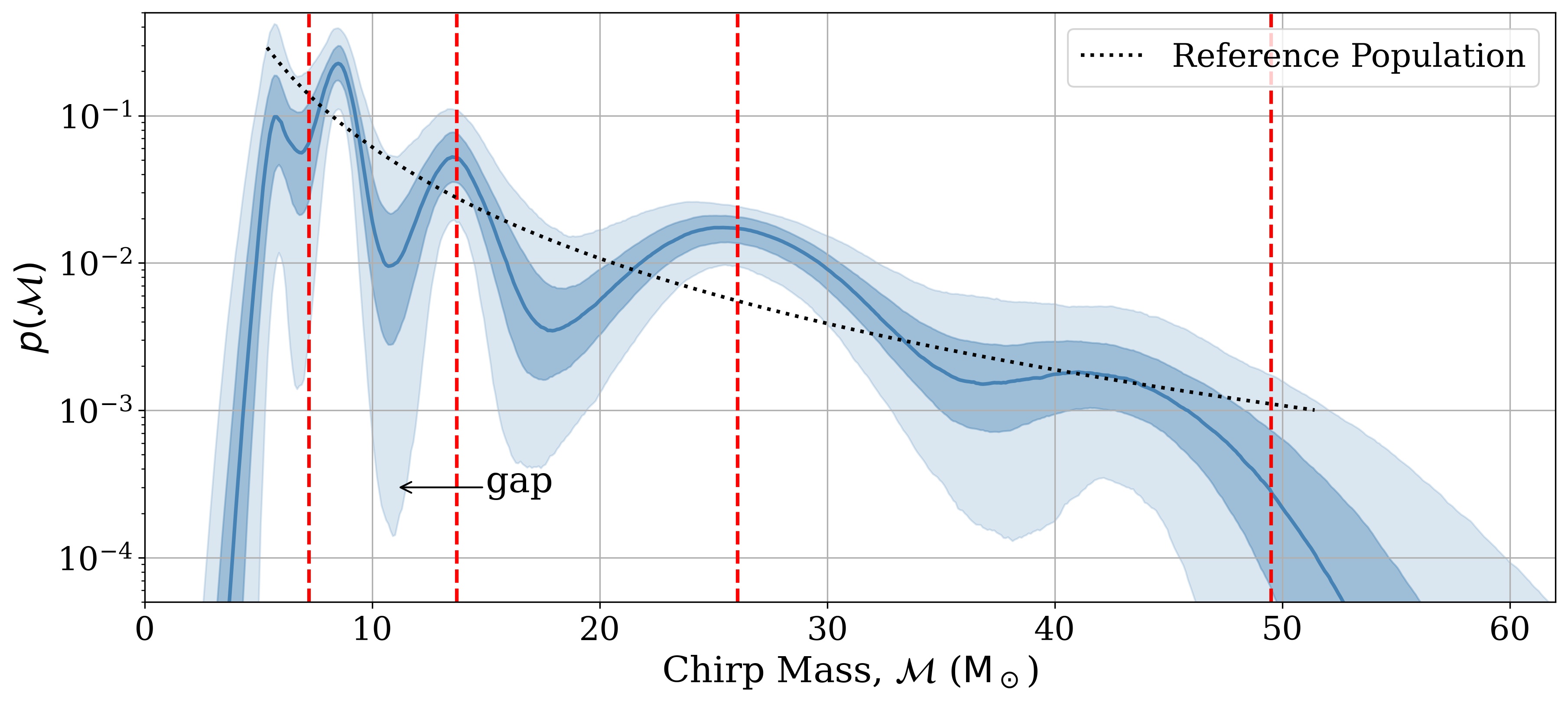}
    \includegraphics[width=0.95\textwidth]{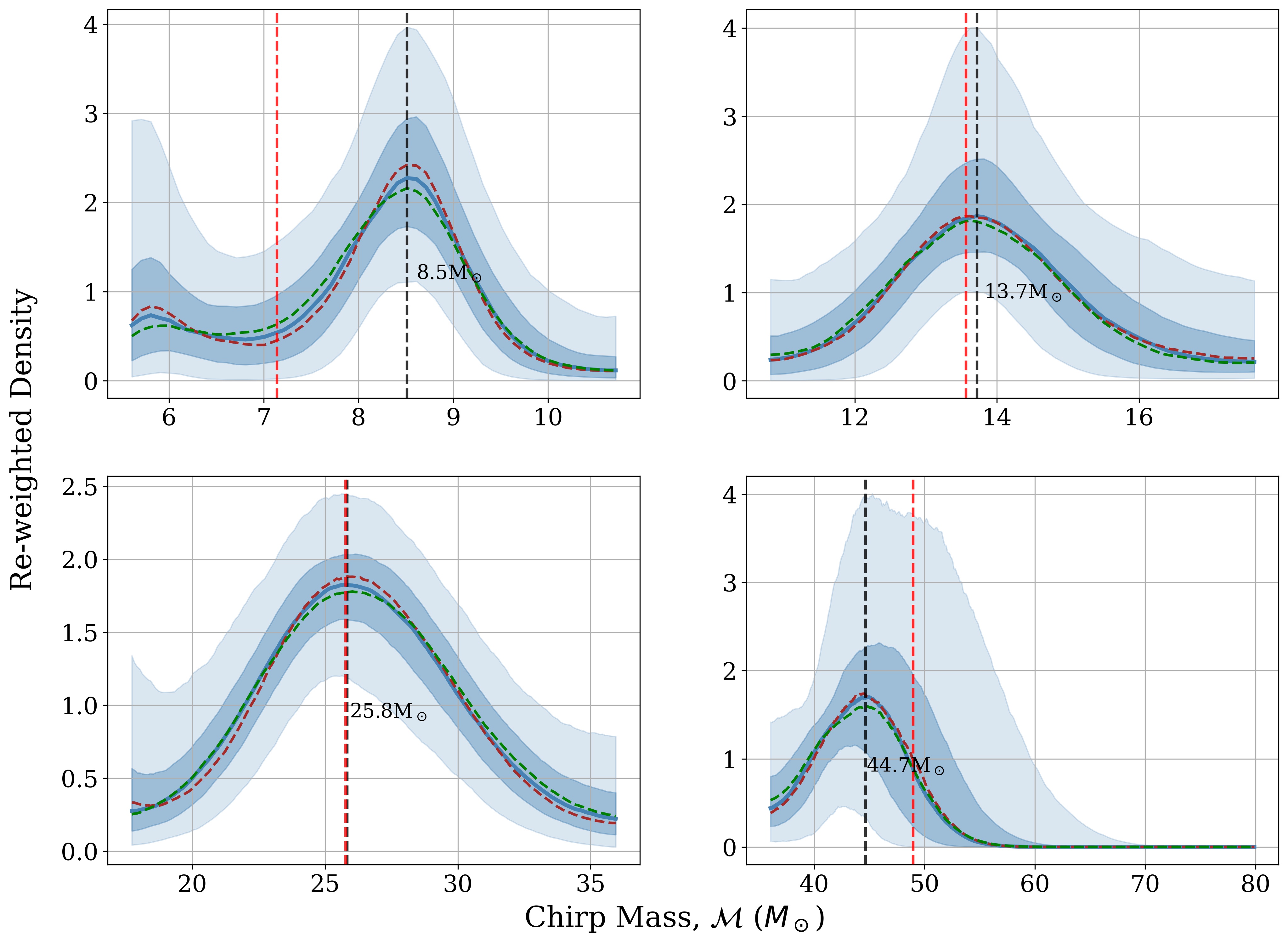}
    \caption{Top) The reconstructed one-dimensional chirp mass distribution. The solid line shows the posterior mean and the shaded regions show the 50\% and 90\% confidence interval. The dotted black curve is the reference power-law model. The chirp mass distribution shows evidence for a number of peaks in the distribution.  The dashed red lines indicate locations of a set of hierarchical peaks, separated by an approximate factor of 1.9, with the position of the first peak at $7.3 M_\odot$. The lack of mergers in the chirp mass range $10-12 M_{\odot}$ is labeled as 'gap'. Bottom) Comparison of the peaks with the best fit power-law in the shown mass range (discussed in detail in Section \ref{sec:hierarchy}). The blue curve is the ratio of reconstructed chirp mass's posterior density with the power-laws and the shaded region are the 50\% and 90\% confidence regions. The dashed black lines indicate the location of the local maxima. For comparison, this figure includes extracted peaks obtained by using 7 (green) and 13 (brown) components in the mixture.
    }
    \label{fig:post_mc}
\end{figure*}

\begin{figure*}
    \centering
    \includegraphics[width=0.95\textwidth]{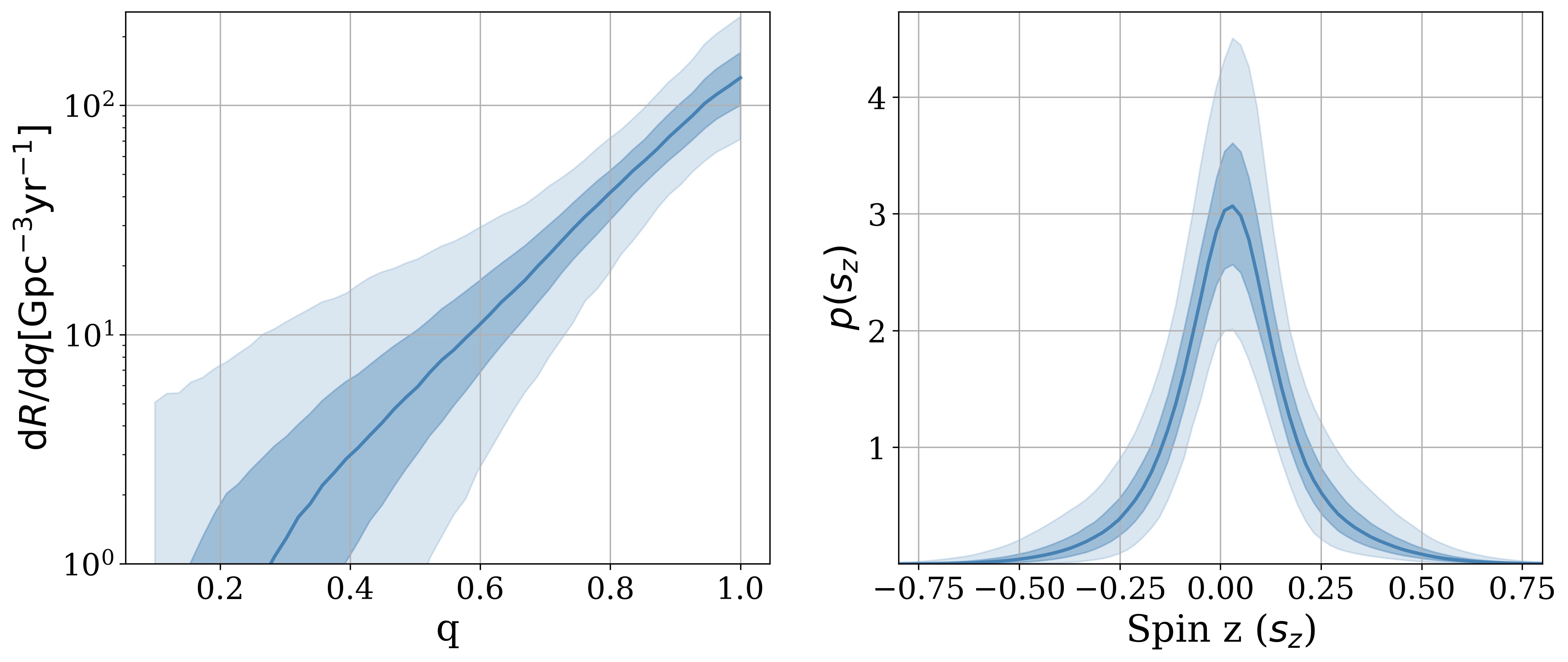}
    \includegraphics[width=0.95\textwidth]{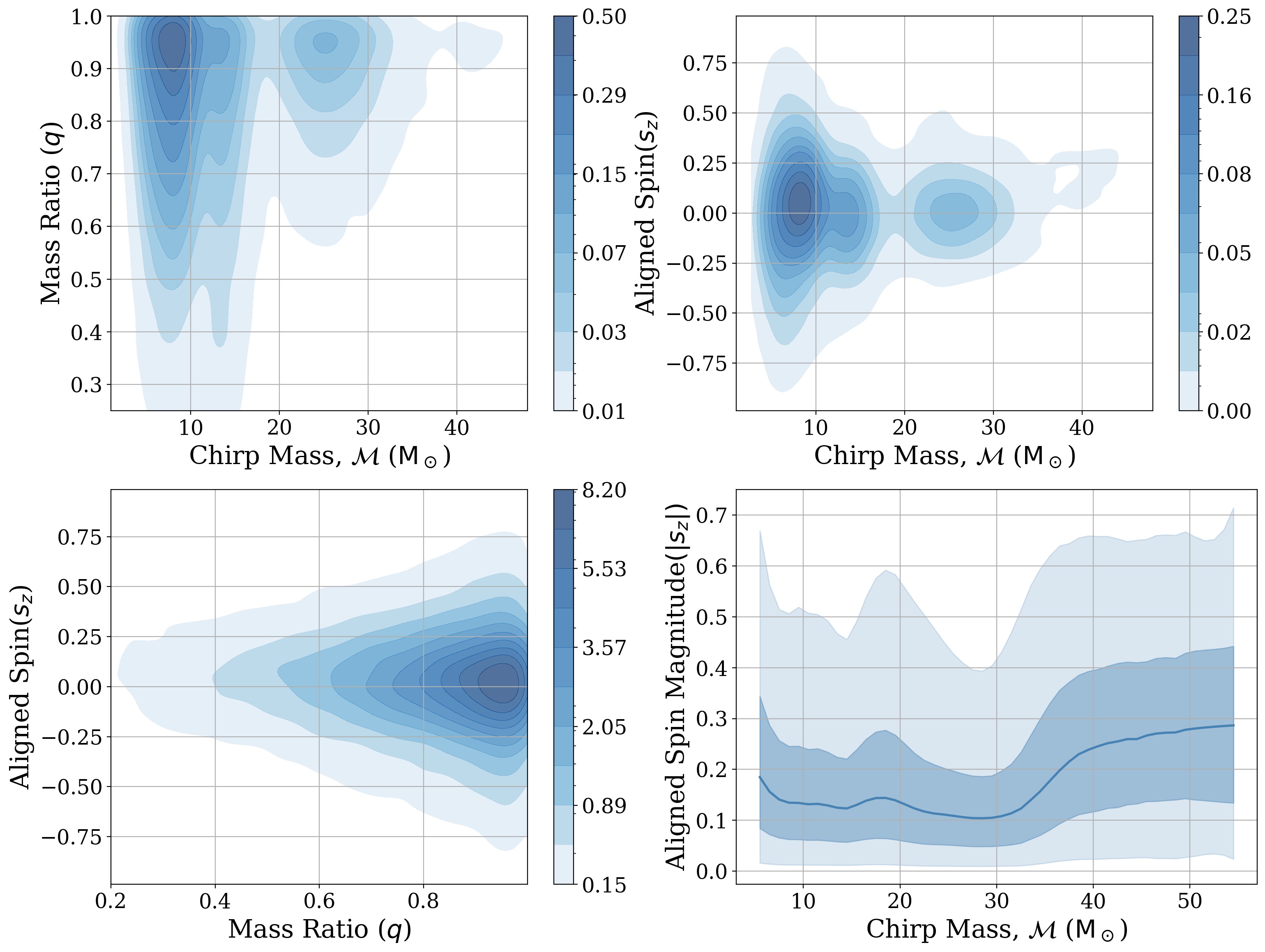}
    \caption{Top) The reconstructed one-dimensional mass ratio and aligned spin distributions. The solid line shows the posterior mean and the shaded regions show the 50\% and the 90\% confidence interval. The mass ratio distribution is clearly peaked towards equal masses, but with support for mass ratios of at least 5:1. The reconstructed aligned spin distribution shows support for small aligned spins, with the distribution, peaked near zero and 90\% of the distribution contained within the range [\lspinz, \rspinz]. 
    Middle and Bottom Left) Contour plot of reconstructed two-dimensional distributions for mass ratio, aligned spin, and the chirp mass. There are ten contours enclosing credible intervals ranging from 5\% to 95\%.  Neither the chirp mass or the aligned spin show a dependence of the mass ratio. Bottom Right) This plot shows the variation of aligned spin magnitude with the chirp mass. The spins magnitude monotonically increases for chirp masses higher than 30$M_\odot$.}
    \label{fig:mchirp_q_sz}
\end{figure*}

The underlying population properties inferred from observations are shown in Figures \ref{fig:post_mc} and \ref{fig:mchirp_q_sz}.  For each of the parameters, we show the 1-dimensional posterior distribution, marginalized over the other dimensions.  We also show the two-dimensional distributions of chirp mass with both mass ratio and aligned spin components.

\subsection{Chirp Mass}

Figure \ref{fig:post_mc} shows the reconstructed chirp mass distribution. Overall, the distribution is weighted towards low masses, with an overall peak at around $\peakonemch M_{\odot}$ and an approximately power-law decay at higher masses.  However, the chirp mass distribution shows a significant structure.  In addition to the primary peak, there are further maxima at $\peaktwomch$ and $\peakthreemch M_{\odot}$, and excess with respect to the power-law model at $\peakfourmch M_{\odot}$.  We observe that the peaks in the chirp-mass distribution occur at approximately a factor of two separations. Figure \ref{fig:post_mc} shows reconstructed chirp mass distribution and the mass normalised distribution that compares them with the best fit power-laws. We also observe that there is a lack of mergers with chirp masses of $10.0-12.0 M_{\odot}$, roughly midway between the first and the second peak.

\begin{figure*}
    \centering
    \includegraphics[width=0.95\textwidth]{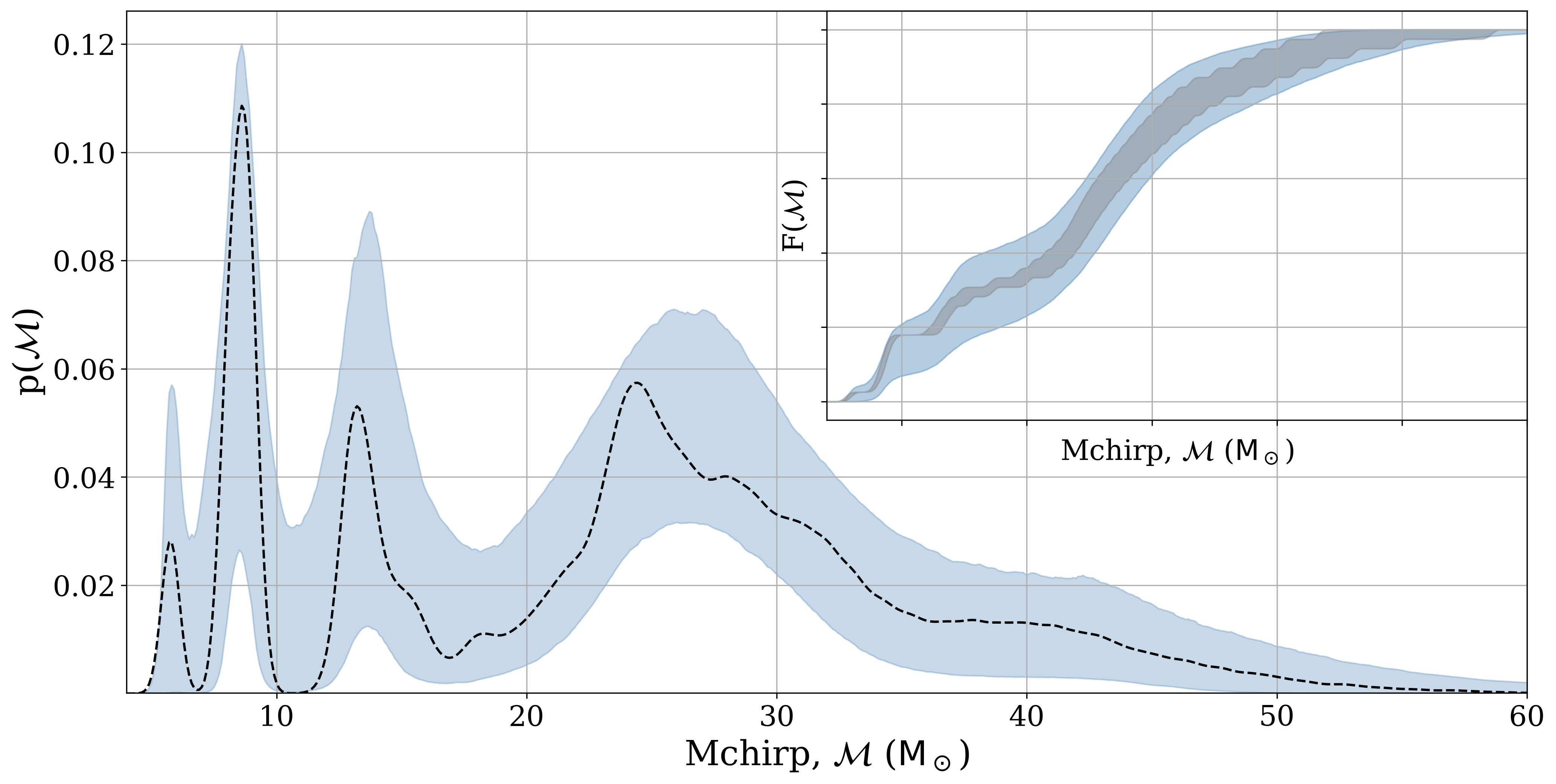}
    \caption{The blue band is the 90\% confidence of the density of the posterior predictive obtained after applying selection effects to the reconstructed chirp mass distribution. The dashed curve is the mean observed chirp mass distribution. Inset: The blue band is the 90\% confidence of the cumulative probability of the posterior predictive obtained after applying selection effects to the reconstructed chirp mass distribution. The grey band is the 90\% confidence obtained by bootstrapping various realisations of the observed data. Each realisation of the observed data is generated by re-weighting the chirp mass estimate of the observations to the reference population and selecting one data point from each one of them.}
    \label{fig:mchirp_scaled_pdf}
\end{figure*}

The observed structure in the inferred chirp mass distribution arises directly from the observed distribution.  The flexibility of VAMANA in modeling complex distributions can be seen from Figure \ref{fig:mchirp_scaled_pdf}. The reconstructed chirp mass distribution embraces the observed chirp mass distribution very well.  The posterior predictive distribution shows that the observed population lies within the 90\% distribution inferred from the population across the whole mass range.

\subsection{Mass ratio and Spins}

The inferred distributions for mass ratio and aligned spin components of the black holes are shown in Figure \ref{fig:mchirp_q_sz}.  The mass ratio is well modeled by a decaying power law, with a peak at equal masses.  The spins are well modeled by a Gaussian distribution, peaked close to zero, although high spins are dis-favoured.  There is little evidence for structure in the mass-ratio and spin distributions.  This is to be expected, as these parameters are less accurately measured than the chirp mass.  Furthermore, the choice of the power-law function in modeling the mass ratio will inadvertently impact the measurement of the spins as the two parameters are significantly correlated \citep{2013PhRvD..87b4035B, 2018ApJ...868..140T}. Due to this correlation, the mass ratio is measured far less accurately, and although the accurate measurement of chirp mass is mostly independent of other parameters for a wide range of masses, the accurate measurement of components masses directly depends on the accurate measurement of mass ratio. Thus, features of astrophysical origin might be more pronounced in the chirp mass distribution compared to the component mass distribution. 

The separate components of VAMANA model different parts of the population in the three-dimensional space of defined chirp mass, mass ratio, and aligned spin. This allows us, in principle, to model variation of the mass ratio or spins with the mass of the binary.  The lower plots in Figure \ref{fig:mchirp_q_sz} show two-dimensional distributions of the mass ratio/spin with chirp mass.  In neither case is there significant evidence for a mass dependence in the observed distribution.  In particular, the reconstructed mass ratio and aligned spin have a 90\% confidence interval of [\lmratio, \rmratio] and [\lspinz, \rspinz] respectively. However, this can be attributed to the dominant contribution from the lower chirp masses which show support for smaller spins. There seems to be a monotonic increase in the spin magnitude for chirp masses greater than 30$M_\odot$

\subsection{Merger Rate}

\begin{figure}
    \centering
    \includegraphics[width=0.5\textwidth]{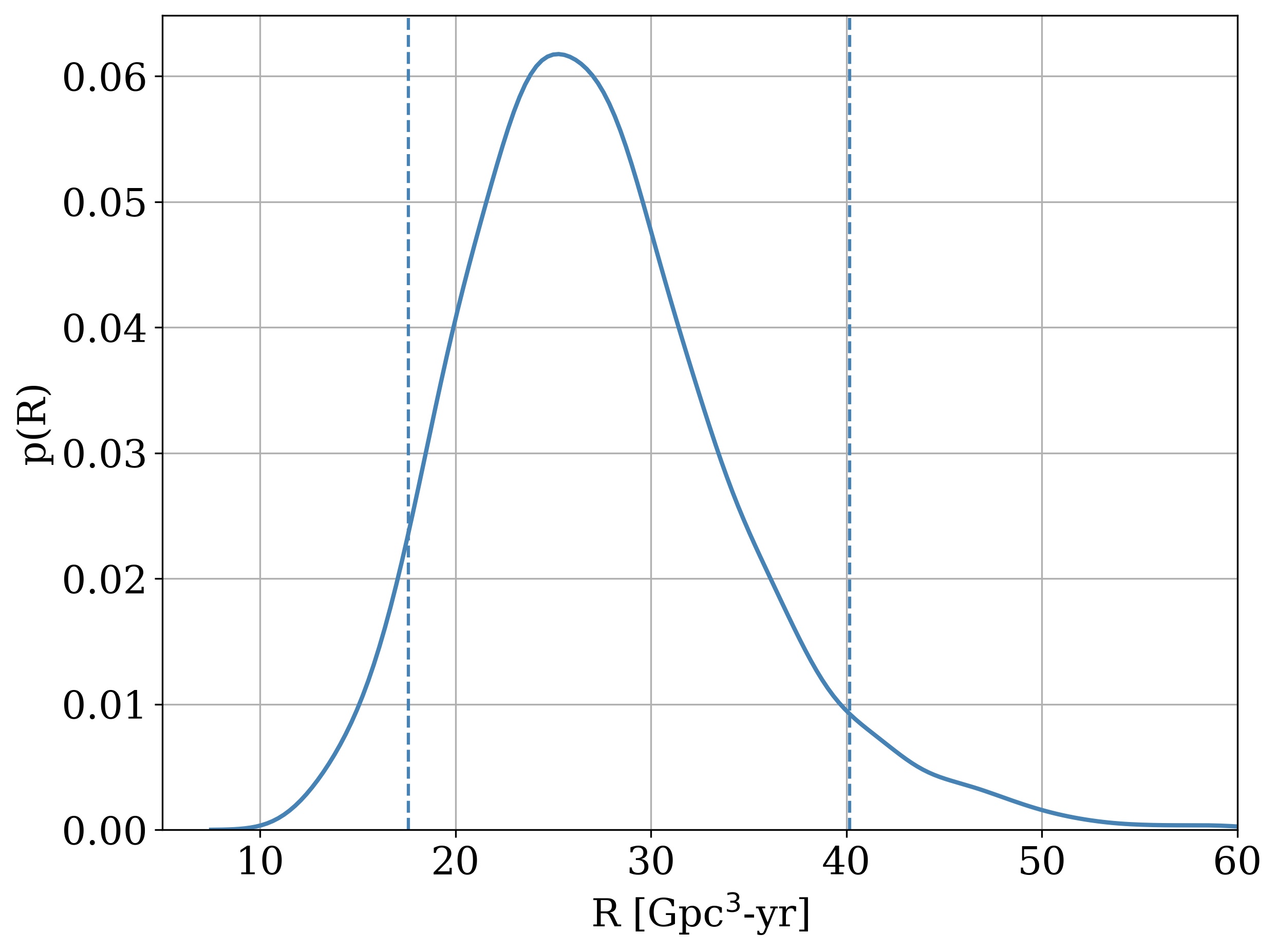}
    \caption{The posterior on the merger rate of binary black-holes for a non-evolving rate in redshift. The dashed lines enclose the 90\% confidence interval which has been estimated to be [\lrate, \rrate] $\mathrm{Gpc}^{-3}\mathrm{yr}^{-1}$.}
    \label{fig:rates}
\end{figure}

Figure \ref{fig:rates} shows the posterior on the merger rate, assuming a rate that is constant in comoving volume.  Certainly, invoking a different redshift evolution, for example, to follow the star formation rate, will impact the inferred merger rate.  For non-evolving merger rate density, we estimate the merger rate to be $\ratem^{+\rateu}_{-\ratel}\,\mathrm{Gpc}^{-3} \mathrm{yr}^{-1}$; the merger rate reported in \citep{2020arXiv201014533T} is $23.9^{+14.9}_{-8.6} \mathrm{Gpc}^{-3}\mathrm{yr}^{-1}$. The two mean values are close, however, the 90\% uncertainty for our result is slightly smaller, likely due to the accurate fitting of the observed chirp mass distribution, as seen in Figure \ref{fig:mchirp_scaled_pdf}.

\section{Structure in the Mass Spectrum}
\label{sec:hierarchy}

Let us now return to consider in detail the observed structure in the mass distribution.  As previously mentioned, there are clusters of events with peaks around \peakonemch, \peaktwomch, \peakthreemch\ and \peakfourmch $M_\odot$.  Each peak occurs at approximately double the mass of the previous peak. In order to investigate the location and significance of these peaks, we plot them relative to the underlying population.  To do so, we split the mass range into four regions, with boundaries at $[5.4, 10.8, 17.7, 37, 70.] M_\odot$ and, for each region, perform a simple power-law fit to the distribution.  In the lower plots in Figure \ref{fig:post_mc}, we show the probability distribution normalized by the best-fit power in each region.  This allows us to clearly identify the peaks. 

\begin{figure*}
    \centering
    \includegraphics[width=0.95\textwidth]{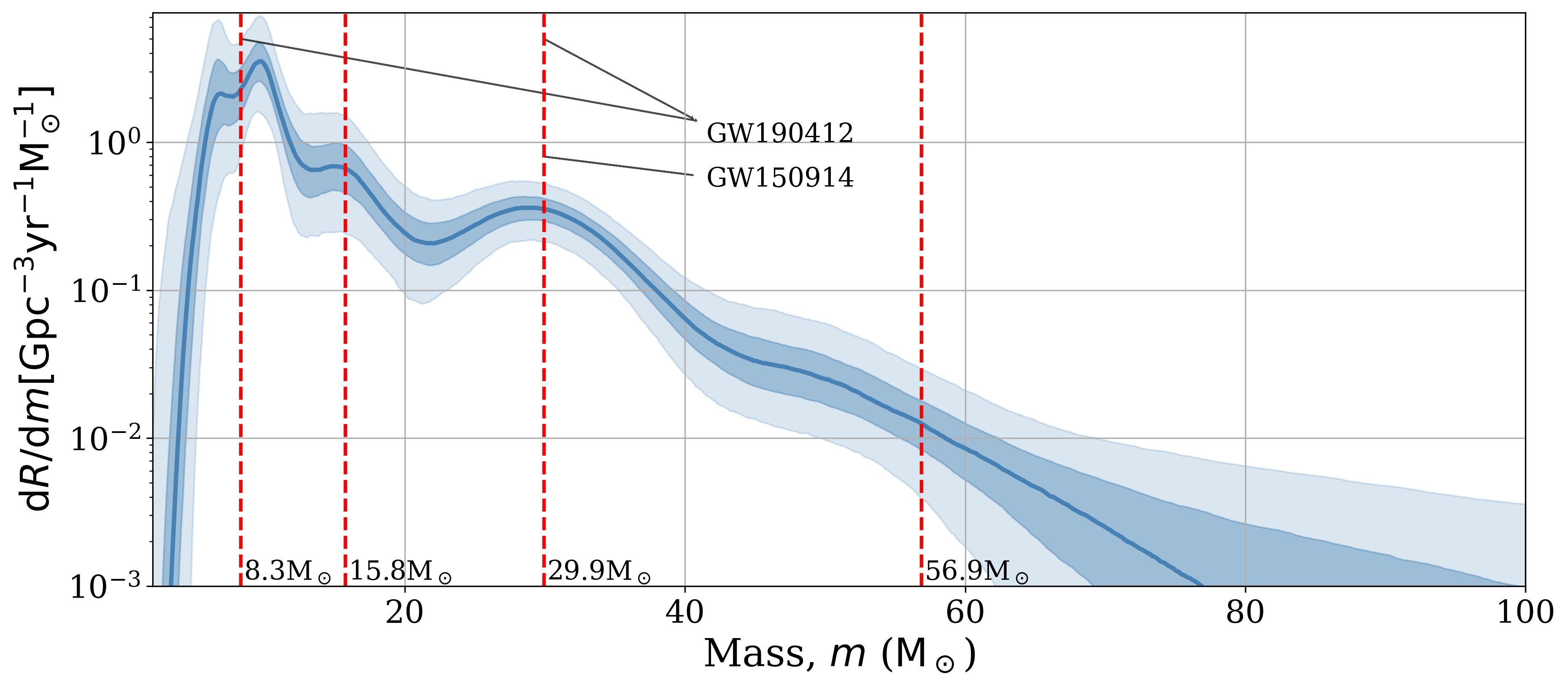}
    \caption{Top) Component mass distribution constructed for binary black holes. The solid line is the mean distribution while the light shaded region shows 90\% confidence interval and the dark shaded region shows 50\% confidence interval. The distribution loosely follows a broken power law but has additional features. The red dashed lines indicate the component mass for equal mass corresponding to the chirp mass values indicated by red lines in figure \ref{fig:post_mc}. We also associate the peaks consistent with the masses of the exceptional events GW150914 and GW190412.}
    \label{fig:component_mass}
\end{figure*}

We can obtain the distribution of component masses from the inferred chirp mass and mass ratio distributions.  
Figure \ref{fig:component_mass} shows the reconstructed component mass distribution. We again identify four peaks in the posterior distribution of the mass spectrum although, due to the poor measurement of mass ratio, these are less clearly defined than in the chirp mass distribution.  The location of these peaks follows directly from the location of peaks in the chirp mass.  For an equal mass binary, $m_{1} = m_{2} = 1.15 \mathcal{M}$, and this sets the location of the peaks in component mass.  However, the fact that the mass ratio is not accurately measured leads to a greater width of the peaks in the component mass distribution.  Here, the mass peaks are observed at \peakonem, \peaktwom, \peakthreem\, and \peakfourm $M_{\odot}$.

The analysis presented in \cite{2020arXiv201014533T}  shows a similar structure in the mass distribution (Figure 4 shows this most clearly), with peaks in the primary mass at around $10, 35$ and $70 M_{\odot}$.  We note, however, that the most flexible mass distribution used models a power law with two peaks, so could not fit a distribution with four peaks.  The fact that the peaks in figure \ref{fig:component_mass} are at different locations is \textit{expected} as we are showing the mass distribution for \textit{both} components rather than just the primary, as reported in \cite{2020arXiv201014533T}.  The relationship between these will depend upon the distribution of mass ratios.  Assuming a mass ratio distribution $\propto q^{1.4}$, as obtained in \cite{2020arXiv201014533T}, the peak of the primary mass distribution will occur at a 20\% greater mass than the peak in the component mass distribution (while the secondary will be at 20\% lower mass).  Thus, the features observed in \cite{2020arXiv201014533T} are consistent with the three observed peaks in component masses of \peakonem, \peakthreem\ and \peakfourm $M_\odot$.  Due to the greater flexibility of our model, and by inferring the better-measured chirp mass distribution, we are able to identify an additional peak at $\peaktwom M_{\odot}$.

The occurrence of these peaks can be attributed to clustering in the observed chirp mass distribution. The first and fourth clusters have a limited number of observations, however, clustering is apparent in the remaining chirp mass range. We have verified that the Akaike information criterion \citep{AIC} is minimised for 3 Gaussians when modeling is performed on the \emph{observed} chirp mass distribution at masses above 10.8 $M_\odot$. The chirp mass structure cannot be confidently explained by a simple power-law model as we calculate a Bayes factor between the mean reconstructed chirp mass distribution and the reference chirp mass distribution to be 60. Moreover, we replace the second peak in the mean reconstructed chirp-mass distribution with a best-fit power-law to approximate the power-law + Gaussian model used in \cite{2020arXiv201014533T}. We compare this new distribution with the mean reconstructed chirp-mass distribution and obtain a Bayes factor of around 8 suggesting the peak at 14 $M_\odot$ is significant. The peaks are complemented by the lack of mergers with chirp masses of $10.0-12.0 M_{\odot}$. The probability that none of the observations are produced within this chirp-mass range by the reference chirp-mass distribution is less than 5\%. 

The mass-gap and peaks can be associated in simplest terms if we allow for the notion that the first peak arises because of a pile-up of black-hole due to a mass gap and the following peaks due to hierarchical mergers \citep{2002MNRAS.330..232C, 2016ApJ...831..187A, 2019MNRAS.486.5008A, 2019PhRvD.100d3027R, 2020arXiv200601867F}. When we account for around 5\% loss in mass expected due to the emission of gravitational waves \citep{PhysRevD.95.064024, Healy:2014yta} the masses of the peaks align very well, as can be seen in Figure \ref{fig:post_mc}. For reference, we show mass locations, each approximately $1.9$ times the previous one. Although the first peak is dominant in the inferred rate, it is actually determined by a smaller number of observations than the second and third peaks, as can be seen from the observed distribution in Figure \ref{fig:mchirp_scaled_pdf}. More observations are needed to accurately reconstruct the first peak. We note that the masses of the binary GW190412 are consistent with the expected location of the first and the third peak, while the masses of the black holes in the first observation, GW150914, are consistent with the location of the third peak. We also note the challenges for a hierarchical model, specifically a) the existence of a mass gap in the $13 M_\odot$ region, b) observed low spins for higher generation mergers in contrast to expectations \citep{2007ApJ...659L...5C, 2007PhRvL..98i1101G, 2008PhRvD..77b6004B}, and c) preferential pairing for equal mass systems (strictly speaking, preferential mass-pairing is only needed for the mergers involving black holes in the first generation; absence of which will fill the $10 - 12 M_\odot$ gap).
In \cite{2020arXiv201105332K}, using a phenomenological mass model, the authors find evidence for hierarchical mergers in the events from GWTC-2. However, their approach and the resulting implication is significantly different from the presented work.

\section{Astrophysical Implications}
\label{hiermerge}

The mass distribution of field binaries is expected to follow a power-law like distribution with the maximum mass of the binaries sensitive to the metallicity and the initial mass function of the progenitor stars. At solar metallicities \citep{2012ApJ...759...52D}, the chirp mass distribution for mergers is predicted to peak in the range $6-8 M_{\odot}$, consistent with the first peak observed here. At lower metallicities, the black hole mass distribution is expected to extend to higher masses. Pair-instability supernovae can impose an upper limit on the maximum mass of the binary as well as introduce a build-up at high masses \citep{1964ApJS....9..201F, 1967ApJ...150..131R, 1984ApJ...280..825B, 2002ApJ...567..532H}. Population synthesis models that simulate complex physics of stellar evolution expect the maximum black-hole to many tens of solar mass \citep{2019MNRAS.485..889S}. However, the results presented here provide initial evidence of a lack of black hole binaries in the chirp mass range 10--12 $M_\odot$.  If supported by future observations, this will become an important piece of information to constrain the stellar evolution physics employed in binary black-hole formation. 

Overall, the reconstructed mass-distribution in Figure \ref{fig:post_mc} and \ref{fig:component_mass} is roughly a broken power-law along with the four peaks and it is possible that more than one formation channel is contributing to the observed black hole mergers. Mergers within star clusters might be a promising channel that explains the existence of structure in the mass-spectrum \citep{1993Natur.364..423S, 2000ApJ...528L..17P, 2015PhRvL.115e1101R, 2016ApJ...831..187A, 2019PhRvD.100d1301G}. This could include their formation in active galactic Nuclei \citep{2017MNRAS.464..946S, 2020A&A...638A.119G, 2019PhRvL.123r1101Y} or formation of binaries due to scattering in galactic cusps \citep{2009MNRAS.395.2127O}. The mass spectrum can potentially inform about the many-body dynamics in the star clusters. Specifically, the relative amplitude and the width of the peaks could provide information about the host environment and if the peaks after the mass gap are exclusively due to hierarchical mergers then the host environment must enable retaining a substantial fraction of merger remnants.

Multiple tests of general relativity have been proposed or performed using \ac{GW}s. These tests usually are performed on individual observations where the consistency of the signal is checked against general relativity. Some of these tests checks for the consistency of the residual data with instrumental noise when the best fit \ac{GW} waveform has been subtracted, consistency among the inspiral, merger, and the merger part of the signal, and consistency of the signal's phase evolution against waveforms that have non-general relativistic effects included parametrically \citep{2010PhRvD..82f4010M, 2013LRR....16....9Y, 2016PhRvL.116v1101A, 2019PhRvL.123a1102A, 2019PhRvD.100j4036A}. Some proposals have extended these tests to included multiple observations by performing a hierarchical analysis \citep{2018CQGra..35a4002G, 2019PhRvD..99l4044Z, 2019PhRvL.123l1101I, 2020arXiv201014529T}. If the separation of peaks is hierarchical in nature, once established, the relative locations will quantify the percentage loss of mass in \ac{GW} due to the merger of black holes and will therefore provide an opportunity to test general relativity predictions of energy emission during the merger. Although the absolute location of the peaks depends on the assumed cosmology, the relative position should remain unchanged within the framework of standard cosmology. Reconstructed features in the mass spectrum do not provide any non-gravitational information and thus cannot be used to estimate the Hubble's constant. But, if a feature can be identified in the source mass-spectrum \citep{2012PhRvL.108i1101M, 2019ApJ...883L..42F} it is conceivable to conduct a combined test of general relativity and cosmology.

\section{Conclusion}

We have presented the inferred mass, mass ratio, and spin distributions of black holes binaries using the observations from GWTC-2 \citep{2020arXiv201014527A}. Most notably, we observe a lack of mergers with chirp masses of $10-12 M_{\odot}$ and four distinct peaks in the mass spectrum.  Each peak occurs at approximately 1.9 times the mass of the previous peak. Our conclusions are currently limited by statistics and we await if they stand the test of time as more events in the \ac{GW} catalog get added. If confirmed by future observations these features have far-reaching implications and will provide deeper insights into the channels responsible for the formation and merger of compact binary black holes and potentially novel tests of general relativity.

\section*{Acknowledgement}
The authors would like to thank Bangalore Sathyaprakash and Fabio Antonini for many useful discussions and to Charlie Hoy for providing detailed comments on the text. This work was supported by the STFC grant ST/L000962/1.

We are grateful for the computational resources provided by  Cardiff  University and funded by an STFC grant supporting UK Involvement in the Operation of Advanced LIGO. We are also grateful for computational resources provided by the Leonard E Parker Center for Gravitation, Cosmology, and Astrophysics at the University of Wisconsin-Milwaukee and supported by National Science Foundation Grants PHY-1626190 and PHY-1700765

This research has made use of data, software and/or web tools obtained from the Gravitational Wave Open Science Center (https://www.gw-openscience.org/), a service of LIGO Laboratory, the LIGO Scientific Collaboration and the Virgo Collaboration. LIGO Laboratory and Advanced LIGO are funded by the United States National Science Foundation (NSF) as well as the Science and Technology Facilities Council (STFC) of the United Kingdom, the Max-Planck-Society (MPS), and the State of Niedersachsen/Germany for support of the construction of Advanced LIGO and construction and operation of the GEO600 detector. Additional support for Advanced LIGO was provided by the Australian Research Council. Virgo is funded, through the European Gravitational Observatory (EGO), by the French Centre National de Recherche Scientifique (CNRS), the Italian Istituto Nazionale della Fisica Nucleare (INFN) and the Dutch Nikhef, with contributions by institutions from Belgium, Germany, Greece, Hungary, Ireland, Japan, Monaco, Poland, Portugal, Spain.

\bibliography{references}

\end{document}